\documentclass[prl,nofootinbib,superscriptaddress,twocolumn]{revtex4}
\usepackage[a4paper,left=1.5cm,right=1.5cm,top=3cm,bottom=3cm]{geometry}

\usepackage{amssymb,amsmath,amsfonts}
\usepackage[dvipsnames]{xcolor}
\usepackage{graphicx}
\usepackage{longtable}
\usepackage{verbatim}
\usepackage{color}

\usepackage{color}
\usepackage{hyperref}
\hypersetup{colorlinks, citecolor=bluscuro, linkcolor=black, urlcolor=bluscuro}
\definecolor{rossos}{cmyk}{0,1,1,0.55}
\definecolor{bluscuro}{rgb}{0.15, 0.2, .85}
\definecolor{bluchiaro}{cmyk}{1,.3,0.,0.1}

\graphicspath{{./Figures/}}

\newcommand{\be}{\begin{equation}}
\newcommand{\ee}{\end{equation}}
\newcommand{\bea}{\begin{eqnarray}}
\newcommand{\eea}{\end{eqnarray}}
\newcommand{\beq}{\begin{equation}}
\newcommand{\eeq}{\end{equation}}
\def\beqa{\begin{eqnarray}}

\def\eeqa{\end{eqnarray}}

\def\lsim{\mathrel{\rlap{\lower4pt\hbox{\hskip0.5pt$\sim$}}
    \raise1pt\hbox{$<$}}}         
\def\gsim{\mathrel{\rlap{\lower4pt\hbox{\hskip0.5pt$\sim$}}
    \raise1pt\hbox{$>$}}}         

\newcommand{\arXiv}[2]{\href{http://arxiv.org/pdf/#1}{{\tt [#2/#1]}}}
\newcommand{\arXivold}[1]{\href{http://arxiv.org/pdf/#1}{{\tt [#1]}}}

\begin{document}

\title{A Cosmological Signature of the Standard Model  Higgs Vacuum Instability:\\
Primordial Black Holes as Dark Matter}
\author{J.~R.~Espinosa }
\address{Institut de F\'{\i}sica d'Altes Energies (IFAE), The Barcelona Institute of Science and Technology (BIST),
Campus UAB, 08193 Bellaterra, Barcelona, Spain}
\address{ICREA, Instituci\'o Catalana de Recerca i Estudis Avan\c{c}ats, \\ 
Passeig de Llu\'{\i}s Companys 23, 08010 Barcelona, Spain}
\author{D.~Racco }
\author{A.~Riotto}
\address{D\'epartement de Physique Th\'eorique and Centre for Astroparticle Physics (CAP), Universit\'e de Gen\`eve, 24 quai E. Ansermet, CH-1211 Geneva, Switzerland}

\date{\today}

\begin{abstract}
\noindent
For the current central values of the Higgs and top masses, the Standard Model Higgs potential develops an instability at a scale of the order of $10^{11}\,$GeV. We show that a cosmological signature of such instability could be dark matter in the form of primordial black holes seeded by Higgs fluctuations during inflation. The existence of dark matter might not require physics beyond the Standard Model.
\end{abstract}

\maketitle


\paragraph{Introduction.}
It has been known for a long time that the Standard Model (SM) Higgs potential develops an instability at large field values \cite{instab2,instab}. 
For the current central values of the Higgs and top masses, the quartic coupling $\lambda$ in the Higgs potential becomes negative for Higgs field values  $\gsim 10^{11}$ GeV,
making our  electroweak vacuum not the one of minimum energy. 
While some take this as motivation for the presence of new physics to change this feature, this is not necessarily a drawback of the SM.
Indeed, our current vacuum is  quite stable against both quantum  tunneling in flat spacetime and thermal fluctuations in the early universe \cite{instab,SSTU}.

The situation is different  during inflation \cite{espinosa}.
If the effective mass of the Higgs is smaller than the Hubble rate $H$ during inflation, quantum 
excitations of the Higgs push the Higgs away from its 
minimum.
The classical  value 
of the Higgs 
randomly walks receiving kicks 
 $\sim \pm (H/2\pi)$ each Hubble time and can surmount the potential barrier and fall deep into the unstable side of the potential \cite{espinosa,cosmo2,Arttu}. At the end of inflation, patches where this happened will be anti-de Sitter regions,  and they  are lethal for our universe as they  grow at the speed of light \cite{tetradis}. 
One can derive  upper bounds on $H$, which depend on the reheating temperature $T_{\rm RH}$ and on the Higgs coupling to the scalar curvature or to the inflaton \cite{tetradis,infdec}. 
 
The upper bound on $H$ depends on $T_{\rm RH}$
because, for sufficiently large values of $T_{\rm RH}$, patches  in which the Higgs field probes the unstable part of the potential  can be recovered thanks to the thermal effects after inflation.
Indeed, the mass squared of the Higgs field receives a positive correction proportional to $T^2$ in such a way that in those would-be dangerous regions the Higgs field can roll back down to the origin and be safe.
 
The physical implications of living in a metastable electroweak  vacuum are fascinating and have far-reaching consequences for  cosmology. This has triggered much activity in a field that involves inflationary dynamics, the physics of preheating, the interplay between Higgs properties and observables of cosmological interest, etc. In spite of this richness, a word of warning is in order: the energy scale of this physics is very high and we have no smoking-gun signature (comparable to proton decay for GUTs) that the electroweak vacuum metastability is actually realized in nature (with the exception of the vacuum decay itself!).

One reasonable question to ask is how can we probe, even if indirectly,  the SM Higgs vacuum instability.     
In this short note  we argue that there might be  a cosmological signature of the SM  vacuum instability:
the very presence of dark matter (DM) in our universe. 
We argue that the origin of DM does not need physics beyond the SM: DM may be due to primordial black holes seeded by the perturbations of the Higgs field generated during the last stages of inflation. The black holes may provide the seeds for structure formation \cite{cg,rev}.

The picture 
is the following. During inflation there are   patches where the Higgs has been pushed by quantum fluctuations beyond the potential barrier and is classically rolling down the slope. 
Higgs fluctuations do not contribute significantly to the total curvature perturbation $\zeta$  which is ultimately responsible for the anisotropies in the Cosmic Microwave Background  (CMB).   Higgs perturbations instead  grow to 
 large values in the last  $e$-folds  of inflation, which are irrelevant for observations in the CMB. When inflation ends and reheating takes place, these regions  
 are   rescued by thermal effects and the  Higgs rolls down to the origin of its potential. At later  times, the Higgs perturbations
reenter inside  the Hubble radius and
they provide high peaks  in the matter power spectrum which give rise to Primordial Black Holes (PBH).
We show that these PBHs can provide the DM we see in the universe today. 

Within a more anthropic attitude, one could say that  the electroweak SM instability is beneficial to our own existence as DM is necessary to form structures. 
In the absence of other DM candidates, the SM would be able to provide  the right DM abundance.  As discussed below, although the parameter choices needed for PBH formation might seem finetuned, they would be motivated anthropically. In particular, this mechanism offers an anthropic explanation of why the electroweak vacuum is metastable (but near-critical, very close to being stable).

\bigskip
\paragraph{The dynamics  during inflation.}
We are agnostic about the details of the model  of inflation and the origin of the curvature perturbation responsible for the CMB anisotropies, which we call $\zeta_{\rm st}$. This  $\zeta_{\rm st}$ might be caused by a single degree of freedom \cite{lrreview} or by another mechanism such as the curvaton \cite{curvaton}. Also, we  take a constant Hubble rate $H$ during inflation and suppose that it ends going through a period of reheating characterised by a reheating temperature $T_{\rm RH}$. Of course, 
one can repeat our calculations for a preferred model of inflation.
We suppose that $H$  is large enough to have allowed the SM Higgs to  randomly walk above the barrier of its potential to probe the potentially dangerous unstable region. As a representative value we take $H\simeq 10^{12}$ GeV.

Despite the Higgs negative potential energy, this region keeps inflating as long as the total vacuum energy during inflation is larger, that is, for 
\be
3H^2 m_{\rm P}^2\gsim \frac{\lambda}{4} h_{\rm c}^4, 
\ee
where $h_{\rm c}$ is the Higgs classical value and $m_{\rm P}=2.4\times 10^{18}$ GeV is the reduced Planck mass.
The equation of motion of the classical value of the SM Higgs  is 
\be
\ddot h_{\rm c}+3 H \dot h_{\rm c}+ V'(h_{\rm c})=0\ ,
\label{a0}
\ee
where, as usual, dots represent time derivatives and primes field derivatives.
For the sake of simplicity, from now on we will approximate the potential as 
\be
V(h_{\rm c})=-\frac{1}{4}\lambda h_{\rm c}^4,
\ee 
with $\lambda>0$ running logarithmically with the field scale.  During inflation, $\lambda$ should in fact be evaluated at a scale $\mu$ given by $\mu^2\simeq h_{\rm c}^2+H^2$ \cite{Arttu}. A typical value (for $h_{\rm c}\gtrsim 10^{12}$ GeV) is $\lambda\simeq 10^{-2}$.
In order to make any prediction deterministic and not subject to probability arguments, we are interested in the regime in which the dynamics of the zero mode  of the Higgs is dominated by the classical motion rather than by the randomness of the fluctuations.  We require therefore that in a Hubble time, $\Delta t=1/H$,  the classical displacement of the Higgs 
\be
\Delta h_{\rm c}\simeq -\frac{V'(h_{\rm c})}{3H^2}\ ,
\ee
 is larger (in absolute value) than the quantum jumps
 \be
 \Delta_{\rm q} h\simeq \pm \left(\frac{H}{2\pi}\right).
 \ee
This implies that, inside the inflating region,  $h_{\rm c}$  must be bounded from below
\be
h_{\rm c}^3 \gsim \frac{3H^3}{2\pi\lambda}.
\label{aa}
\ee
We call $t_*$ the initial time at which the Higgs starts its classical evolution.
In this estimate we assume that the motion of the Higgs is friction dominated, that is $\ddot h_{\rm c}\lsim 3 H \dot h_{\rm c}$.
This is true as long as $h_{\rm c}^2\lsim 3 H^2/\lambda$.
If so, the Higgs is slowly moving for a sufficient number of $e$-folds.
The evolution of the classical value of the Higgs is
\be
\label{pp}
h_{\rm c}(N)\simeq \frac{h_{\rm e}}{\left(1+2 \lambda h_{\rm e}^2 N/3H^2\right)^{1/2}},
\ee
where  we have introduced the number of $e$-folds till the end of inflation $N$ and denoted by $h_{\rm e}$ the value of the classical Higgs field at the end of inflation.

Meanwhile, Higgs fluctuations are generated.  Perturbing around the slowly-rolling classical value of the Higgs field and accounting for metric perturbations as well, the Fourier transform of the perturbations of the Higgs field satisfy the equation of motion (in the flat gauge)
\be
\label{a}
\delta\ddot{h}_{k}+3 H\delta\dot{h}_{k}+\frac{k^2}{a^2}\delta  h_{k}+V''(h_{\rm c})\delta  h_{k}=\frac{\delta  h_{k}}{a^3m_{\rm P}^2}\frac{{\rm d}}{{\rm d}t}\left(\frac{a^3}{H}\dot{h}_{\rm c}^2\right) ,
\ee
where $a$ is the scale factor and the last term accounts for the backreaction of the metric perturbations. 
 Driven by the Higgs background evolution in the last $e$-folds of inflation, the Higgs perturbations grow significantly after leaving the Hubble radius. The reason is the following. Having numerically checked that the last term in Eq.~(\ref{a}) is negligible, the Higgs perturbations and $\dot h_{\rm c}$ solve the same equation  on scales larger than the Hubble radius  $k\ll aH$, as can be seen by taking the time derivative of Eq.~(\ref{a0}). Therefore the two quantities must be proportional to each other during the evolution  and on super-Hubble scales
\be
\delta  h_{k}=C(k)\,\dot{h}_{\rm c}(t).
\ee
Matching at Hubble crossing $k=aH$ this  super-Hubble solution for $\delta  h_{k}$ with its standard wave counterpart on sub-Hubble scales  implies that 
\be
C(k)=\frac{H}{\dot{h}_{\rm c}(t_k)\sqrt{2 k^3}},
\ee
where $t_k$ is the time when the mode with wavelength $1/k$ leaves the Hubble radius. The growth of $\delta h_k$ is therefore dictated by the
growth of $\dot{h}_{\rm c}$. These Higgs perturbations will be responsible for the formation of PBHs. In fact, we should deal with the  comoving curvature perturbation $\zeta$ which is gauge-invariant and reads (still in the flat gauge)
\be
\zeta=H\frac{\delta\rho}{\dot \rho}=\frac{\dot{\rho}_{\rm st}}{\dot\rho}\zeta_{\rm st}+\frac{\dot{\rho}_h}{\dot\rho}\zeta_h=\frac{\dot{\rho}_{\rm st}}{\dot\rho}\zeta_{\rm st}+H\frac{\delta{\rho}_h}{\dot\rho},
\ee
where  $\zeta_h$ is the Higgs perturbation.  
We assume $\zeta_{\rm st}$ is conserved during inflation on super-Hubble scales and, for simplicity,  that there is no energy transfer with Higgs fluctuations. In the curvaton model, for instance,   $\zeta_{\rm st}$ could be even zero on large scales during inflation.

 Using Eqs. (\ref{a0}) and (\ref{a})  (again with the negligible last term dropped), one then obtains
 \be
  \delta\rho_h(k\ll aH) = \dot{h}_{\rm c}\delta \dot h_k+V'(h_{\rm c})\delta h_k =
 -3HC(k)\dot{h}^2_{\rm c}.
 \ee
 Since  $ \dot{\rho}_h=\dot{h}_{\rm c}(\ddot{h}_{\rm c}+V'(h_{\rm c}))=-3H\dot{h}^2_{\rm c},$ 
  one can easily show (and we have checked it numerically) that during inflation
and on super-Hubble scales $\zeta_h$ reaches the plateau
\be
\zeta_h(k\ll aH) = H\frac{\delta\rho_h}{\dot{\rho}_h} = H C(k)
 = \frac{H^2}{\sqrt{2 k^3}\dot{h}_{\rm c}(t_k)}.
\ee
This is the quantity which gives the largest contribution to $\zeta$ in the last few $e$-folds
before the end of inflation.

\bigskip
\paragraph{Dynamics after inflation: reheating.}
At the end of inflation, the vacuum energy which has driven inflation gets converted into thermal relativistic degrees of freedom, a process dubbed reheating.
For simplicity, we  suppose that this conversion  is instantaneous, in such a way that the reheating  temperature is 
$T_{\rm RH}\simeq 0.5\cdot(H\, m_{\rm P})^{1/2}$,
obtained by energy conservation and  taking the number of relativistic degrees of freedom to be about  $10^2$.  For our representative value of $H=10^{12}$ GeV, 
we obtain $T_{\rm RH}\simeq 10^{15}$ GeV.
Due to the thermal effects, the Higgs potential receives thermal corrections such that the potential is quickly augmented by the term
\cite{tetradis}
\be
V_T \simeq \frac{1}{2}m^2_T  h_{\rm c}^2,\,\,m_T^2\simeq 0.12\, T^2\,e^{-h_{\rm c}/(2\pi T)} ,
\ee
(a fit that is accurate for $h\lesssim 10\, T$ in the region of interest and includes the effect of ring resummation).
If the maximum temperature  is larger than the value of the Higgs $h_{\rm e}$ at the end of inflation, or more precisely if $T^2_{\rm RH }\gsim \lambda h_{\rm e}^2$,
the corresponding patch is thermally rescued and 
the initial value of the Higgs immediately after  the end of inflation coincides with  $h_{\rm e}$.
The classical value of the Higgs field starts oscillating around the origin, see Fig.~1.  The  Higgs fluctuations oscillate as well  with the average value remaining 
constant
and the amplitude  slowly increasing for a fraction of $e$-folds. 
  At the same time, the curvature perturbation, with power spectrum  
${\cal P}_{\zeta}=k^3/(2\pi^2)\left|\zeta_k\right|^2$,
given in Fig.~2, gets the largest contribution from the Higgs fluctuations. After inflation, the long wavelength Higgs perturbations 
 decay after several oscillations into radiation curvature perturbation which, being radiation now the only component, will stay constant on super-Hubble scales.  We have taken the  Higgs damping rate  to be $\gamma_h\sim  3g^2 T^2/(256\pi m_T)\sim 10^{-3} T$ \cite{ee} (where $g$ is the SU(2)$_L$ coupling constant). This value has been derived by noticing that for a thermal Higgs mass
$m_T\simeq 0.34\, T$, the one-loop absorption and direct decay channels for quarks and gauge bosons are forbidden, and the damping occurs through two-loop diagrams involving gauge bosons. Therefore, we have  evaluated the value of the curvature perturbation after a fraction of $e$-fold.

\begin{figure}[t!]
\includegraphics[width=\columnwidth]{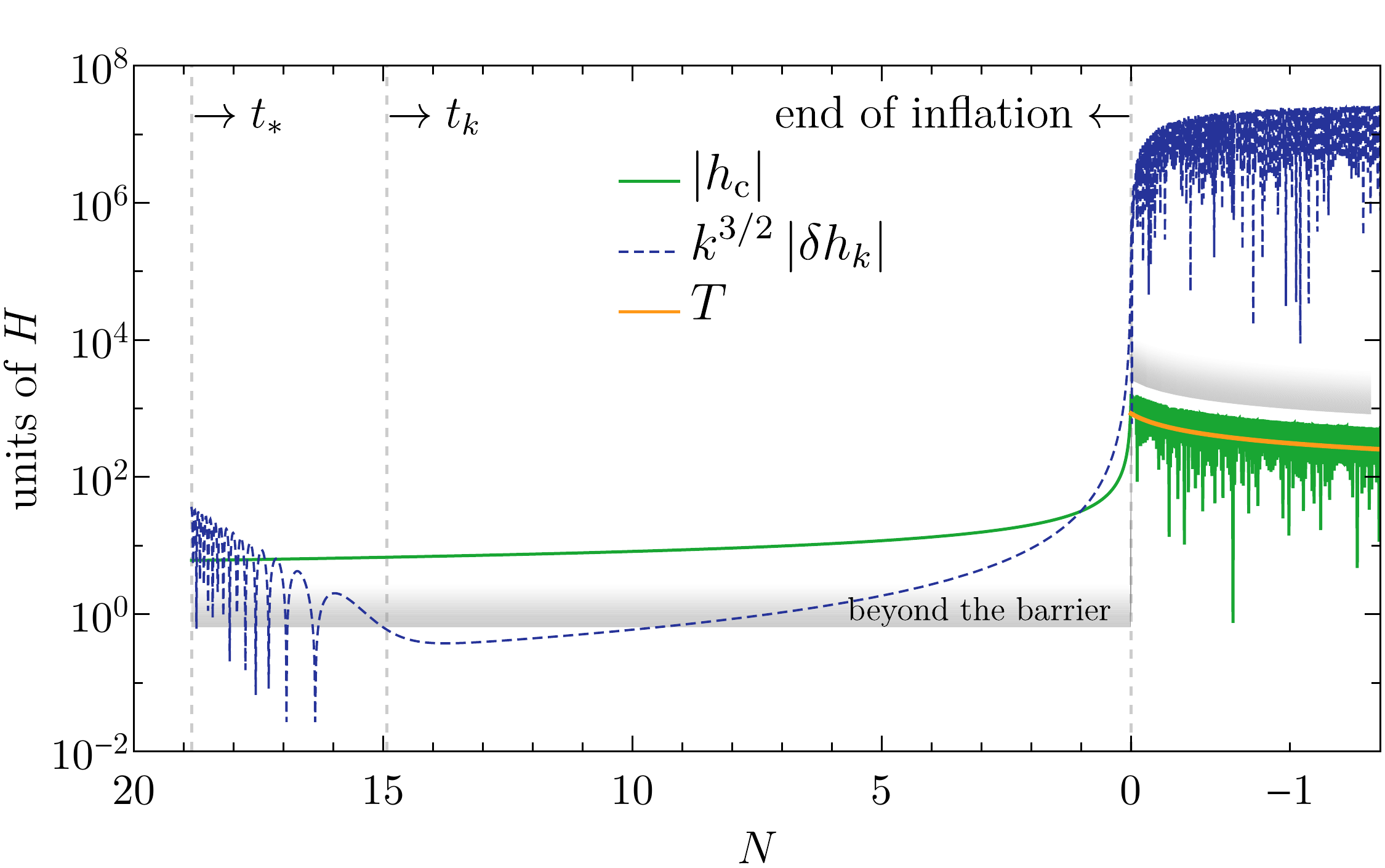}
\caption{Evolution of $H$, $T$, $h_{\rm c}$, $\delta h_k$ during the last $e$-folds of inflation, for $k=50\, a(t_*)H$ where $t_*$ is defined to be the time  when  $h_{\rm c}$ starts its classical evolution. The region of $h_{\rm c}$ beyond the top of the potential barrier is shaded gray.
}
\label{fig: evolution}
\end{figure}

\begin{figure}[t!]
\includegraphics[width=\columnwidth]{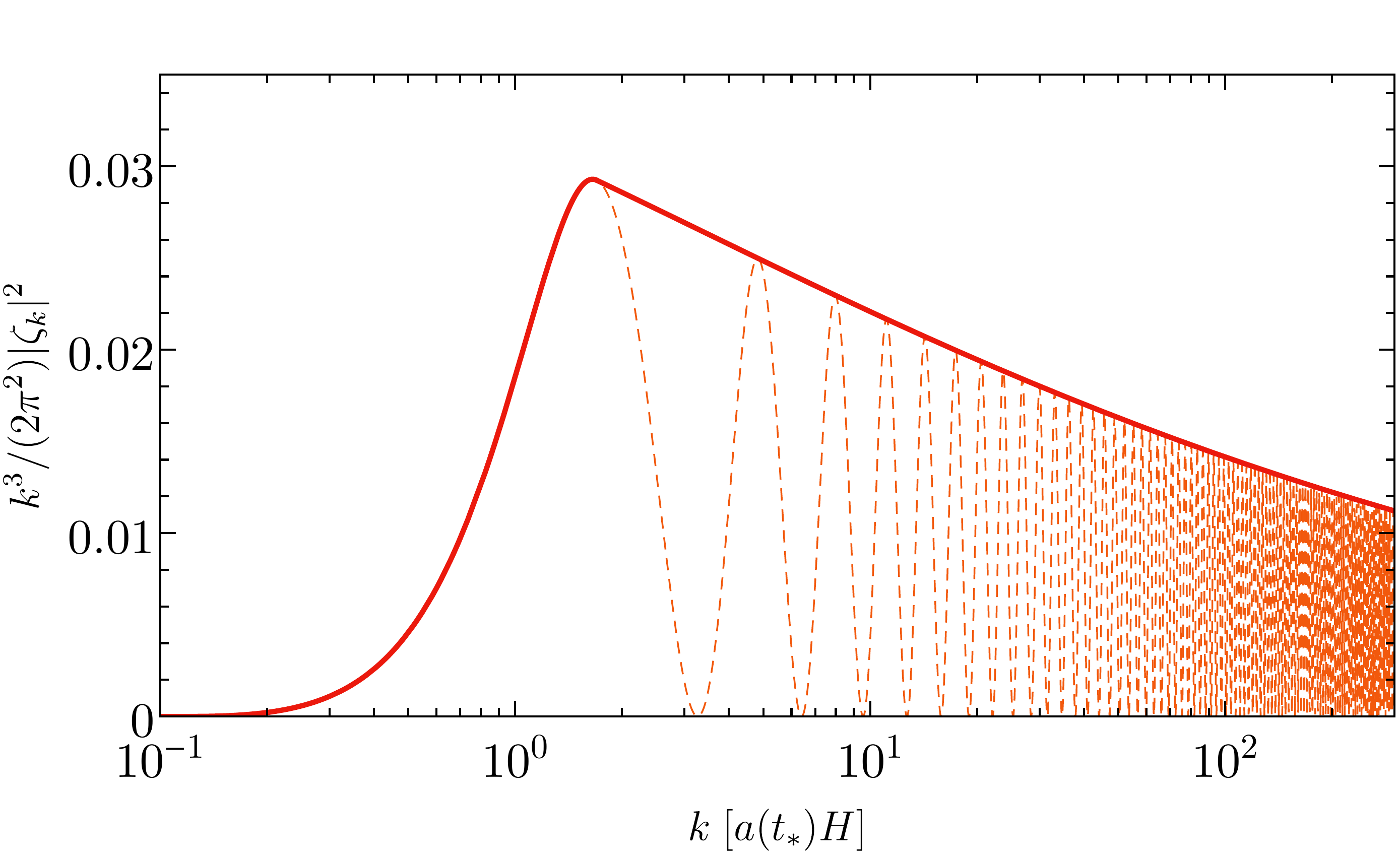}
\caption{The power spectrum $\mathcal P_\zeta$, shown as the envelope of different modes (averaged over their uncorrelated time phases).}
\label{fig: power spectrum}
\end{figure}

\bigskip
\paragraph{Generation of Primordial Black Holes.} 
After inflation, the Hubble radius  grows and the perturbations generated during the last $e$-folds of inflation are the first  to reenter the horizon. If they are large enough, they  collapse to form  PBHs almost immediately after horizon reentry,  see Ref.~\cite{Carr} and references therein. 
A    given region  collapses to  a PBH if  the density contrast  (during the  radiation era) $\Delta(\vec x)=(4/9a^2H^2)\nabla^2\zeta(\vec x)$ is above a critical value $\Delta_{\rm c}$. Typically $\Delta_{\rm c} \sim 0.45$   \cite{c}.
As a result,  in order to obtain a significant number of PBHs, the   power spectrum on small scales must be sizeable.
The mass of a PBH at formation and corresponding to the density fluctuation leaving the Hubble radius $N$ $e$-folds
before the end of inflation is about \cite{jn}
$ M\simeq\frac{m_{\rm P}^2}{H}e^{2N}$.
We first define the variance of the density contrast 
\be
\sigma^2_\Delta(M) =\int_0^\infty {\rm d}\ln k \,W^2(k,R){\cal P}_{\Delta}(k),
\ee
where $W(k,R)$ is a gaussian window function  smoothing out the density contrast on the comoving horizon length  $R\sim 1/aH$. 
The mass fraction $\beta(M)$ of the universe which ends up into PBHs at the time of formation $t_M$  is 
\be
\label{ll}
\beta(M)=\int_{\Delta_{\rm c}}^\infty \frac{{\rm d}\Delta}{\sqrt{2\pi}\,\sigma_\Delta}e^{-\Delta^2/2\sigma_\Delta^2}\simeq\frac{\sigma_\Delta}{\Delta_{\rm c}\sqrt{2\pi}} e^{-\Delta_{\rm c}^2/2\sigma^2_\Delta},
\ee
The total contribution of  PBHs at radiation-matter equality ($t_{\rm eq}$) is obtained  by integrating the  fraction $\beta(M,t_{\rm eq})=a(t_{\rm eq})/a(t_M)\beta(M)$ \cite{cg} 
\be
\Omega_{\rm PBH}(t_{\rm eq})=\int_{M_{\rm ev}(t_{\rm eq})}^{M(t_{\rm eq})}{\rm d}\ln M\, \beta(M,t_{\rm eq}), 
\ee
where $M_{\rm ev}(t_{\rm eq})\simeq 10^{-21} M_\odot$  is the lower mass  which has survived evaporation at equality
and $M(t_{\rm eq})$ is the horizon mass at equality (which for our purposes can be taken equal to infinity).

\begin{figure}[t!]
\includegraphics[width=\columnwidth]{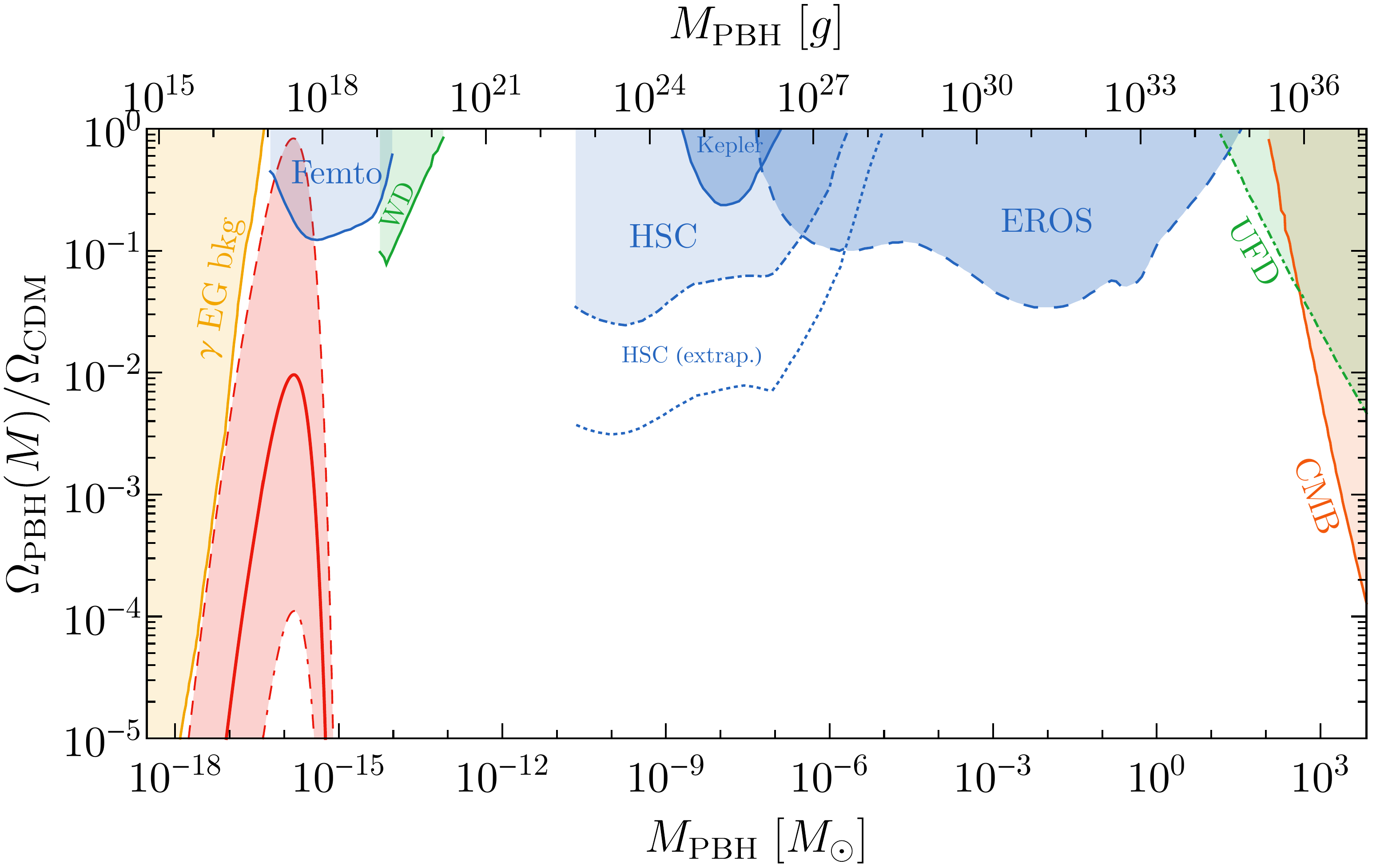}
\caption{Spectrum of PBHs at formation generated by the mechanism we discuss (solid red refers to $S_3=0$, and dashed lines to $S_3=\pm 1$), superimposed with the experimental constraints on monocromatic PBH spectra (from Ref.~\cite{japanese} and references therein): in yellow, the observations of extra-galactic $\gamma$-ray background; in blue, femto-, micro- and milli- lensing observations from Fermi, Eros, Kepler, Subaru HSC; in green, dynamical constraints from White Dwarves and Ultra-Faint Dwarf galaxies; in orange, constraints from the CMB.}
\label{fig: Omega PBH}
\end{figure}

Fig.~\ref{fig: Omega PBH} shows the resulting mass spectrum of PBHs at their formation time. 
The position of the peak in the PBH mass spectrum is set by the mode $k_*$  that exits the Hubble radius during inflation when the Higgs zero mode starts its classical evolution.
To be on the safe side we ask  that the interesting range of PBH masses  is  large enough to avoid the bounds from evaporating PBHs by  the present time. This requires   the dynamics to last about 17  $e$-folds before the Higgs hits the pole in Eq. (\ref{pp}).  Interestingly this can be achieved in the SM for realistic values of the Higgs and top masses and  $\alpha_s$: In our numerical example we use  $M_h=125.09$ GeV, $M_t=172$ GeV, and $\alpha_s=0.1184$.

In our findings we have not included the fact that the mass of the PBH is not precisely the mass contained in the corresponding horizon volume, but in fact obeys a  scaling relation with  initial perturbations \cite{jn} or the fact that the  threshold is shape-dependent \cite{asp}. 
Furthermore, we have not accounted for the fact that the threshold amplitude and the final black hole mass depend on the initial density  profile of the perturbation \cite{misao}. 
We estimate that the first two effects change the abundance by order unity. 
The third effect would require a thorough study of the spatial correlation of density fluctuations. 
Nevertheless, we have included in Fig.~\ref{fig: Omega PBH} the possible effect of non-Gaussianity in the PBH mass function. 
To estimate the impact of non-Gaussianity is not an easy task, as one needs to evaluate the second-order contribution to the comoving curvature perturbation $\zeta_2$. 
A rough estimate based on Ref. \cite{wands} gives $\zeta_2={\cal O}(1)\zeta_1^2$ and therefore we include in Fig.~\ref{fig: Omega PBH} two bands corresponding to $S_3=\pm 1$, where $S_3=\langle \Delta^3\rangle/\sigma_\Delta^4$ is the skewness which appears in the modification of the arguments of the exponential in Eq.~\eqref{ll} via the shift $\nu^2\rightarrow \nu^2 \left[1-S_3\frac{\sigma_\Delta}{3}\left(\nu-2-\frac{1}{\nu^2}\right)\right]$, with $\nu=\Delta_{\rm c}/\sigma_\Delta$ \cite{mr}. The shift in the final abundance is not negligible, but we stress that there will be a set of parameters in our model which can provide the right final abundance. 
We also stress that the primordial abundance of PBHs depends in a very sensitive way on the value of $t_*$, keeping fixed all the other parameters. 
This does not come as a surprise as the function $\beta(M)$ is exponentially sensitive to $\nu$. 
In this sense the anthropic argument based on the necessity of having DM would justify a tuned initial PBH abundance. As a final warning, one should keep in mind that splitting the metric
into  background and perturbations might be questionable for large
perturbations.

From the time of equality to now, the PBH mass distribution will slide  to larger masses  due to merging. While the  final word can  only be said through N-body simulations, one can expect merging to shift  the spectrum to higher masses even by orders of magnitude \cite{cc} and  to spread the spectrum, but maintaining  the abundance. Accretion, on the other hand, increases both the masses and the abundance of PBHs as DM. On the other side, both merging and accretion  help to render the PBHs more long-living. 
To roughly account for  an increase of the current abundance by a representative factor $10^2$ because of accretion,  we have properly set the abundance at formation time to be
$\Omega_{\rm PBH}/\Omega_{\rm DM}\sim 10^{-2}$ (higher values can be achieved). 
It would be certainly interesting to analyse these issues in more detail and account for the fact that the abundance of PBH has to be of the right magnitude during standard Big Bang Nucleosynthesis.

\bigskip
\paragraph{Conclusions.}
If the scenario we have presented were in fact realized in nature, we can highlight three points as 
the most relevant. First, the SM would be capable of explaining DM by itself (supplemented by a period of inflation  that is well motivated by other reasons). This has a double side: the SM provides a DM candidate in the form of PBHs and also provides the mechanism necessary to create the PBH seeds during inflation via the quantum fluctuations of the Higgs field in the unstable part of the Higgs potential. Both aspects (DM candidate and PBH generation mechanism) go against the common lore that physics beyond the SM  is needed. In fact, if this scenario were correct, the Higgs field would not only be responsible for the masses of elementary particles but also for the DM content of our universe. Second, the PBH generation mechanism gives an anthropic handle on Higgs near-criticality which would be explained as needed to get sufficient DM so that large enough structures can grow in the universe. Finally, the PBHs responsible for DM would represent a conspicuous cosmological signature of the actual existence of an unstable range in the Higgs potential at large field values.

\bigskip
\paragraph{Some extra considerations about the Non-Gaussianity.}
In this subsection, added in v2 of the paper, we offer some considerations about the non-Gaussianity of the perturbations.

To evaluate the non-Gaussianity  at the instant at which the perturbations re-enter the Hubble radius we  proceed as follows. 
During the radiation phase, we have
\be
\rho_h=\rho_h+\delta\rho_{h,1}+\frac{1}{2}\delta\rho_{h,2}=m_T^2h_{\rm c}^2+2 m_T^2h_{\rm c}\delta h_1+m_T^2 \delta h_1^2,
\ee
so that 
\be
\frac{\delta\rho_{h,2}}{\rho_h}=\frac{1}{2}\left(\frac{\delta\rho_{h,1}}{\rho_h}\right)^2=
8\zeta_{h,1}^2,
\ee
where we have used again the fact that (in the flat gauge)
\be
-\zeta_{h,1}=H\frac{\delta\rho_{h,1}}{\dot{\rho}_h}=H\frac{\delta\rho_{h,1}}{-4H\rho_h}=-\frac{1}{4}\frac{\delta\rho_{h,1}}{\rho_h}.
\ee
The total gauge-invariant curvature perturbation at second-order is \cite{mw,uss,uss1,noi} 
\begin{multline}
-\zeta_{2}=\psi_2-2\frac{\delta\rho_{h,1}}{\dot{\rho}}(\psi_1+2H\dot{\psi}_1)
+H\frac{\delta\rho_{h,2}}{\dot{\rho}}-2\frac{H}{\dot{\rho}^2}\delta\dot{\rho}_{h,1}\delta\rho_{h,1}\\
+H^2\frac{(\delta\rho_{h,1})^2}{\dot{\rho}^2}\left(
\frac{\ddot{\rho}}{H\dot{\rho}}-\frac{\dot H}{H^2}-2\right),
\end{multline}
where we have assumed that on small scales only the perturbation of the Higgs field is relevant. Defining $r_h=\dot{\rho}_h/\dot\rho$ and using the
fact that during the radiation phase $\dot H=-2H^2$, $\dot\rho=-4H\rho$ and $\ddot\rho=-6H\dot\rho$, we find
(using again the flat gauge)
\be
-\zeta_{2} 
= -2r_h(1-r_h)\zeta_{h,1}^2=-2\frac{1}{r_h}(1-r_h)\zeta_{1}^2,
\label{so}
\ee
where we have used the relation $\delta\dot{\rho}_{h,1}=-4H\delta{\rho}_{h,1}$. 
A similar computation gives
\be
-\zeta_{2,h}=\left(-2+8-6\right)\zeta_{h,1}^2=0
\ee
and therefore the Higgs perturbation is Gaussian. This is important for what comes later on.

One can ask about the non-Gaussianity during inflation. Writing $h=h_{\rm c}+ \delta h_1+\delta h_2/2$, the equation for $\delta h_2$ on super-Hubble scales is
\be
\delta\ddot{h}_2+3H\delta\ddot{h}_2 +V''\delta h_2+ V'''\left(\delta h_1\right)^2=0,
\ee
from which one deduces that, if $\delta h_1(t,\vec x)=C(\vec x)\dot{h}_{\rm c}(t)$, then 
\be
\delta h_2(t,\vec x)=C^2(\vec x)\ddot{h}_{\rm c}(t).
\ee
During inflation the gauge-invariant second-order Higgs curvature perturbation is
\begin{multline}
-\zeta_{h,2}=\psi_2-2\frac{\delta h_{1}}{\dot{h}_{\rm c}}(\psi_1+2H\dot{\psi}_1)
+H\frac{\delta h_{2}}{\dot{h}_{\rm c}}\\
-2\frac{H}{\dot{h}_{\rm c}^2}\delta\dot{h}_{1}\delta h_{1} 
+H^2\frac{(\delta h_{1})^2}{\dot{h}_{\rm c}^2}\left(
\frac{\ddot{h}_{\rm c}}{H\dot{h}_{\rm c}}-\frac{\dot H}{H^2}-2\right),
\end{multline}
In the flat gauge one finds
\be
-\zeta_{h,2}=-2\zeta_{h,1}^2\,\,\,\,(\textrm{during inflation}).
\ee
Another way of finding the result (\ref{so}) is the following. In the absence of interactions, the Higgs and radiation have a conserved curvature perturbation \cite{con}
\be
\zeta_i(\vec x)=-\psi(t,\vec x)+\frac{1}{3}\int^{\rho_i(t,{\vec x})}_{\rho_i(t)}\frac{{\rm d}\tilde{\rho}_i}{\tilde{\rho}_i+\tilde{P}(\tilde{\rho}_i)},\,\,\,\,(i=\gamma,h).
\ee
Assuming that the  Higgs  decays on a uniform (total) density hypersurface corresponding to $\gamma_h=H$, being $\gamma_h$ the decay rate of the Higgs. On this hypersurface one has
\be
\rho_\gamma(t_{\rm dec},\vec x)+\rho_h(t_{\rm dec},\vec x)=\rho(t_{\rm dec}).
\ee
On this hypersurface, we have  have $\zeta= -\psi$.
On the other hand, the  local Higgs  and radiation densities on such decay surface are inhomogeneous
\begin{eqnarray}
\zeta_\gamma&=&\zeta+\frac{1}{4}\ln\frac{\rho_\gamma(t,{\vec x})}{\rho_\gamma(t)},\nonumber\\
\zeta_h&=&\zeta+\frac{1}{4}\ln\frac{\rho_h(t,{\vec x})}{\rho_h(t)},
\end{eqnarray}
and therefore
\begin{eqnarray}
\rho_\gamma(t,{\vec x})&=&\rho_\gamma(t)e^{-4(\zeta-\zeta_\gamma)},\nonumber\\
\rho_h(t,{\vec x})&=&\rho_h(t)e^{-4(\zeta-\zeta_h)}.
\end{eqnarray}
Since the  total density is uniform on the decay surface one finds
\be
(1-r_h)e^{-4\zeta}+r_h e^{-4(\zeta-\zeta_h)}=1,
\ee
where we have assumed that on small scales $\zeta_\gamma=0$. Solving for $\zeta$ one finds
\be
\label{nm}
\zeta_\pm=\pm\frac{1}{4}\ln\left(1-r_h+r_h e^{4\zeta_h}\right).
\ee
In practice, the solution corresponding to $\zeta_{-}$ can be disregarded as one is interested in large values of $\zeta$ when dealing with the primordial black holes. 
Expanding at first order $\zeta_{+}$ one finds $\zeta=r_h\zeta_h$ and at second-order one recovers the relation (\ref{so}). 

Now, the relation (\ref{nm}) allows to compute the non-perturbative probability function for the quantity $\zeta$, by using the relation $P(\zeta_{+}){\rm d}\zeta_{+}=P(\zeta_h){\rm d}\zeta_h$. 
One can first find $P(\zeta_h)$ and then integrate it from a critical value 
\be
\zeta_h(\zeta_{\rm c})=\frac{1}{4}\ln\left(\frac{r_h-1+e^{4\zeta_{\rm c}}}{r_h}\right)
\ee
in order to find the mass fraction of the  primordial black holes at formation time. Typical values in the literature for  $\zeta_{\rm c}$ go from 0.1 to 1.3 \cite{bb}.
The fact that $P(\zeta_h)$ is  Gaussian considerably  simplifies the computation:  the primordial mass fraction $\beta_{\rm prim}(M)$ of the universe occupied by  primordial black holes formed at the time  $t_M$  is therefore given by
\begin{eqnarray}
\label{a}
P(\zeta>\zeta_{\rm c})&=&\beta_{\rm prim} (M)=\int_{\zeta_{\rm c}} {\rm d}\zeta\,P(\zeta)\nonumber\\
&=&\int_{\zeta_h(\zeta_{\rm c})} \frac{{\rm d}\zeta_{h}}{\sqrt{2\pi}\,\sigma_{\zeta_h}}e^{-\zeta_{h}^2/2\sigma^2_{\zeta_h}}\nonumber\\
&=&\int_{\zeta_h(\zeta_{\rm c})} \frac{{\rm d}\zeta_{1}}{\sqrt{2\pi}\,\sigma_{\zeta_1}}e^{-\zeta_{1}^2/2\sigma^2_{\zeta_1}}\,,
\end{eqnarray}
where we have used $\zeta_1=r_h\zeta_h$.
For $\zeta_{\rm c}\simeq 0.5$ and $r_h\simeq 0.01$, one finds the new threshold to be $\zeta_h(\zeta_{\rm c})\simeq 1.6$, which seems to signal that non-Gaussianity makes more difficult
to produce PBHs. We write ``seems" because it is by now accepted in the literature that $\zeta(\vec x)$ is not the best variable to describe the  PBH mass fraction at
formation \cite{bb}. The  density contrast $\Delta(\vec x)$ is  more suitable. This however makes more difficult
to gauge the importance of the non-Gaussianity due to the presence of the Laplacian operator.  One might   evaluate the density contrast at Hubble crossing, so that  $\Delta(\vec x)\simeq(4/9a^2H^2)\nabla^2\zeta(\vec x)\simeq 4/9 \zeta(\vec x)$ and then use the relation among $\zeta(\vec x)$ and the Gaussian $\zeta_h(\vec x)$. Another approach might be
to compute the Laplacian  identifying the PBHs with the peaks of the distribution and therefore dropping the gradients of the fields. 

Notice that one should also include another source of  non-Gaussianity coming from the non-linear mapping between $h(t_{\rm RH})$ and $h(t_{\rm e})$.
This certainly calls for a more thorough analysis to asses the impact of the non-Gaussianity onto the PBH mass distribution.

\bigskip
\begin{acknowledgments}
\paragraph{Acknowledgments.}
We thank G.F. Giudice for comments.  A.R. and D.R. are supported by the Swiss National Science Foundation (SNSF), project {\sl Investigating the Nature of Dark Matter} (no. 200020-159223).
The work of J.R.E. has been partly supported by the ERC
grant 669668 -- NEO-NAT -- ERC-AdG-2014, the Spanish Ministry MINECO under grants  2016-78022-P and
FPA2014-55613-P, the Severo Ochoa excellence program of MINECO (grant SEV-2016-0588) and by the Generalitat grant 2014-SGR-1450. J.R.E. also thanks IFT-UAM/CSIC for hospitality and partial financial support during the final stages of this work.
\end{acknowledgments}

\end{document}